\documentclass{emulateapj}

\slugcomment{Accepted in ApJ}

\shorttitle{ARCADE 2 Instrument}
\shortauthors{Singal et al.}

\begin{document}

\title{THE ARCADE 2 INSTRUMENT \\}

\author{J. Singal\altaffilmark{1}, D.J. Fixsen\altaffilmark{2}$^,$\altaffilmark{3}, A. Kogut\altaffilmark{3}, S. Levin\altaffilmark{4}, M. Limon\altaffilmark{5}, P. Lubin\altaffilmark{6}, P. Mirel\altaffilmark{7}$^,$\altaffilmark{3}, M. Seiffert\altaffilmark{4}, T. Villela\altaffilmark{8},\\ E. Wollack\altaffilmark{3}, and C.A. Wuensche\altaffilmark{8}}


\altaffiltext{1}{Kavli Institute for Particle Astrophysics and Cosmology\\SLAC National Accelerator Laboratory, Stanford University
\\Menlo Park, CA 94025\\
email: jsingal@stanford.edu}
\altaffiltext{2}{University of Maryland}
\altaffiltext{3}{Code 665, NASA Goddard Space Flight Center\\Greenbelt, MD 20771}
\altaffiltext{4}{Jet Propulsion Laboratory\\4800 Oak Drive, Pasadena, CA, 91109}
\altaffiltext{5}{Columbia University\\1027 Pupin Hall, Box 47, New York , NY 10027}
\altaffiltext{6}{University of California, Santa Barbara\\Santa Barbara, CA, 93106}
\altaffiltext{7}{Wyle Information Systems}
\altaffiltext{8}{Instituto Nacional de Pesquisas Espaciais, Divis\~ao de Astrof\'{\i}sica, Caixa Postal 515, 12245-970 - S\~ao Jos\'e dos Campos, SP, Brazil}

\begin{abstract}
The second generation Absolute Radiometer for Cosmology, Astrophysics, and Diffuse Emission (ARCADE 2) instrument is a balloon-borne experiment to measure the radiometric temperature of the cosmic microwave background and Galactic and extra-Galactic emission at six frequencies from 3 to 90 GHz.  ARCADE 2 utilizes a double-nulled design where emission from the sky is compared to that from an external cryogenic full-aperture blackbody calibrator by cryogenic switching radiometers containing internal blackbody reference loads.  In order to further minimize sources of systematic error, ARCADE 2 features a cold fully open aperture with all radiometrically active components maintained at near 2.7 K without windows or other warm objects, achieved through a novel thermal design.  We discuss the design and performance of the ARCADE 2 instrument in its 2005 and 2006 flights.
\end{abstract}

\keywords{instrumentation: detectors -- cosmic microwave background -- radio continuum: galaxies}

\section{Introduction}

ARCADE 2 is part of a long term effort to characterize the cosmic microwave background (CMB) and Galactic and extra-Galactic microwave emission at cm wavelengths.  The background spectrum has been shown to be a nearly ideal blackbody from $\sim$60 to $\sim$600 GHz with a temperature of 2.725 $\pm$.001 K \citep{F96}.  At lower frequencies, however, where deviations in the spectrum are expected, existing measurements have uncertainties ranging from 10 mK at 10 GHz to 140 mK at 2 GHz (see \citet{F04} for a recent review).  These uncertainties are primarily the result of instrumentation systematics, with all previous measurement programs below 60 GHz needing significant corrections for instrument emission, atmospheric emission, or both.  \citet{MW70}, \citet{JW87}, and \citet{S96} are examples of notable CMB absolute temperature measurements from high altitude balloons.  

To improve existing radiometric temperature measurements, an instrument must be fully cryogenic, so that microwave emission from front end components of the instrument is negligible.  In any absolute radiometric temperature measurement, radiation from the source being measured is compared by the radiometer to that from a blackbody emitter of known temperature.  Because of drifts in the gain of the radiometer, a comparison within the radiometer to another blackbody emitter of known temperature is necessary.  Therefore, in order to achieve significantly lower uncertainties than previous measurements, ARCADE 2 is a high altitude balloon-based double-nulled instrument with open-aperture cryogenic optics mounted at the top of an open bucket liquid helium dewar.

A previous generation instrument with observing channels at 10 and 30 GHz, built to test the cold open aperture design, observed in 2003 \citep{F04} and is described by \citet{KI04}.  Following verification of the instrument concepts, the full six frequency ARCADE 2 instrument, with observing channels ranging from 3 to 90 GHz was designed and built, and observed in 2005 and 2006.  Scientific analysis of the 2005 and 2006 flights is presented by  \citet{SR06}, and \citet{F08}, \citet{K08}, and \citet{S08} respectively.  

This paper describes the design and performance of ARCADE 2.  We present the radiometric and thermal properties of the instrument necessary to achieve the results presented in the companion papers, and also describe elements of engineering design that may be useful.

\section{Instrument Design}

ARCADE 2 reduces systematic errors through a combination of radiometer design and thermal engineering.  The instrument core is contained within a large (1.5 m diameter, 2.4 m tall) open bucket liquid helium dewar.  Maintaining such a large volume and mass at cryogenic temperatures in an open environment without significant atmospheric condensation presents considerable instrumental challenges.  The external calibrator, aperture, antennas, and radiometers are maintained at temperatures near 2.7 K through the use of liquid helium tanks fed from the helium bath at the bottom of the dewar by a network of superfluid pumps.  Boiloff helium gas is used for the initial cool-down of components on ascent, and directed in flight to discourage the condensation of ambient nitrogen on the aperture.

\begin{figure}
\includegraphics[width=3.5in]{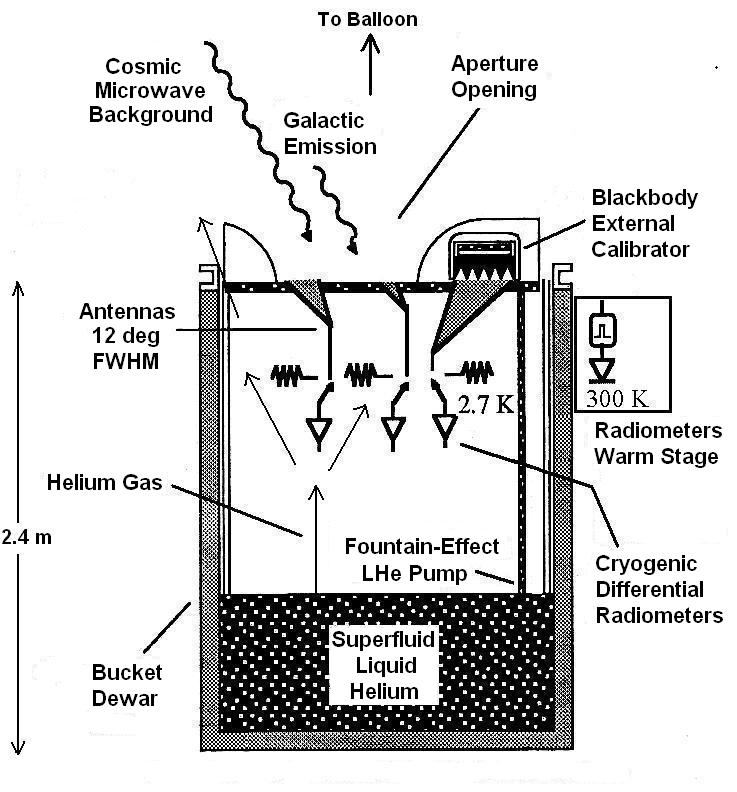}
 \caption{ARCADE 2 instrument schematic, components not to scale.  Cryogenic radiometers compare the sky to an external blackbody calibrator.  The antennas and external calibrator are maintained near 2.7 K at the mouth of an open bucket dewar; there are no windows or other warm objects between the antenna and the sky.  Cold temperatures are maintained at the top of the dewar via boil-off helium gas and tanks filled with liquid helium fed by superfluid pumps in the bath.  For observing the sky, everything shown is suspended below a high altitude balloon.}
\label{schematic}
\end{figure}

Figure \ref{schematic} shows an overview of the ARCADE 2 instrument.  The corrugated horn antennas for each frequency band hang from a flat horizontal aluminum aperture plate at the top of the open dewar.  There are seven observing channels, one each at 3, 5, 8, 10, 30, and 90 GHz, and an additional channel at 30 GHz with a much narrower antenna beam to provide a cross-check on emission in the antenna sidelobes.  Cryogenic temperatures within the external calibrator and internal reference loads, as well as on components throughout the instrument core, are read with ruthenium oxide resistance thermometers.  

A carousel structure containing both a port for sky viewing and the external calibrator sits atop the aperture plate and turns about a central axis to alternately expose the horns to either the sky or the calibrator.  The open port hole and external calibrator are both ellipses measuring 700 mm x 610 mm.  Each radiometer measures the difference in emitted power between radiation incident on the horn and that from an internal blackbody reference load.  The experiment performs a doubly nulled measurement, with the radiometric temperature of the sky compared to the physical temperature of the external calibrator in order to eliminate systematic effects within the radiometer to first order.  The radiometer itself differences the internal reference load from the horn signal allowing a determination of the coupling of radiometer output to instrument temperatures and a near-nulling of the radiometer output to reduce the effects of gain fluctuations.

The ARCADE 2 instrument is flown on a high altitude 790 Ml balloon to the upper atmosphere (37 km altitude) in order to reduce atmospheric emission and contamination from terrestrial microwave sources to negligible levels.  Balloon launch and recovery operations are handled by the Columbia Scientific Balloon Facility (CSBF) in Palestine, TX (31.8\degr\ lat, -95.7\degr\ long.).  

\subsection{Antenna aperture and carousel configuration\label{appconfig}} 

\begin{figure}
 \includegraphics[width=3.5in]{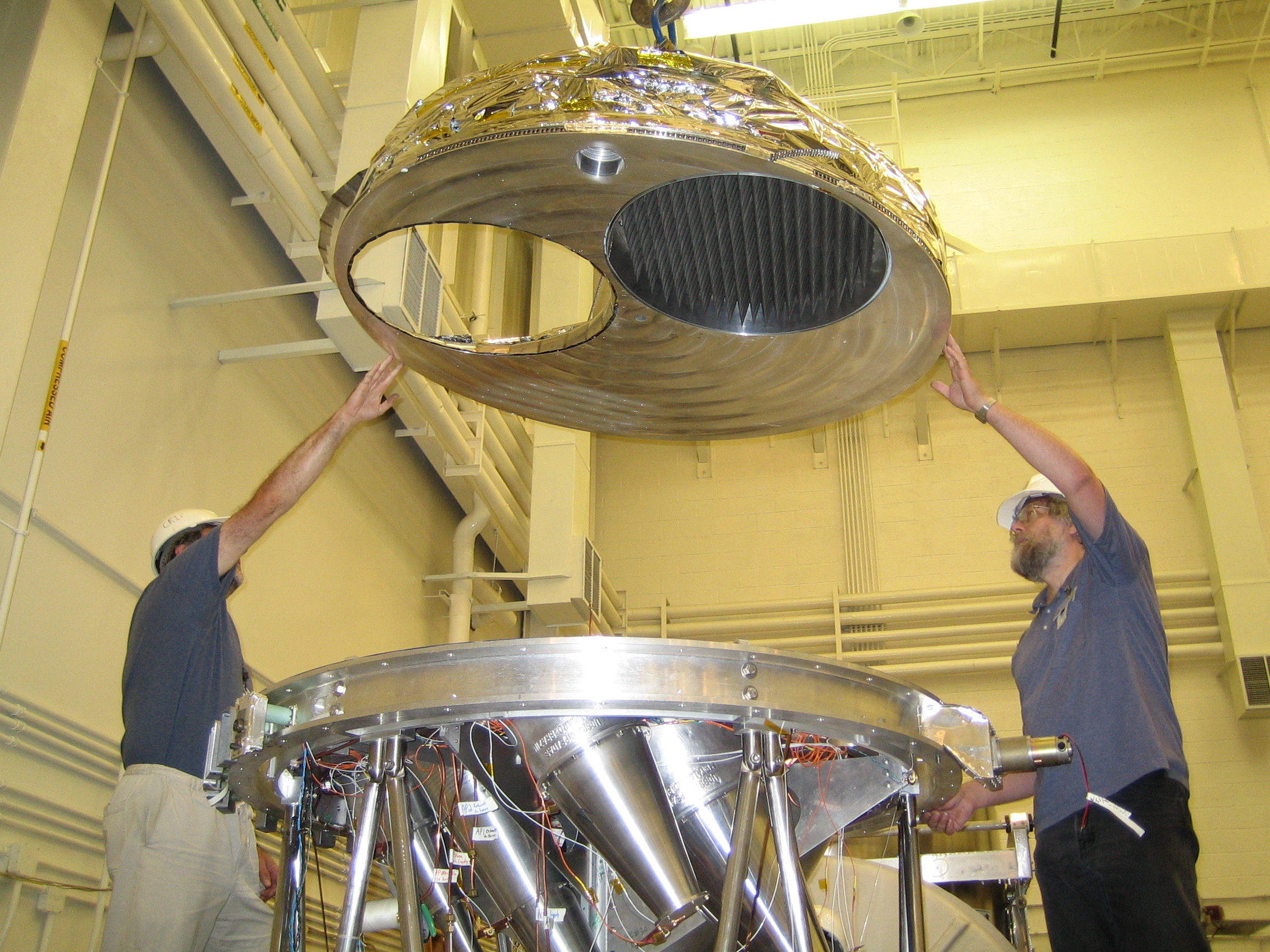}
 \caption{Photograph of the carousel being lowered on top of the aperture plate by two of the authors, showing the port hole for sky viewing and the external calibrator.  The carousel turns atop the aperture plate to expose groups of horns to the sky or to the external calibrator.  The external calibrator is a blackbody emitter consisting of 298 cones of Steelcast absorber cast onto aluminum cores, and the radiometric side is visible in the photograph.   Horn antennas in the core are visible hanging down from the aperture plate.  }
 \label{carousel}
 \end{figure}

\begin{figure}
\includegraphics[width=1.5in]{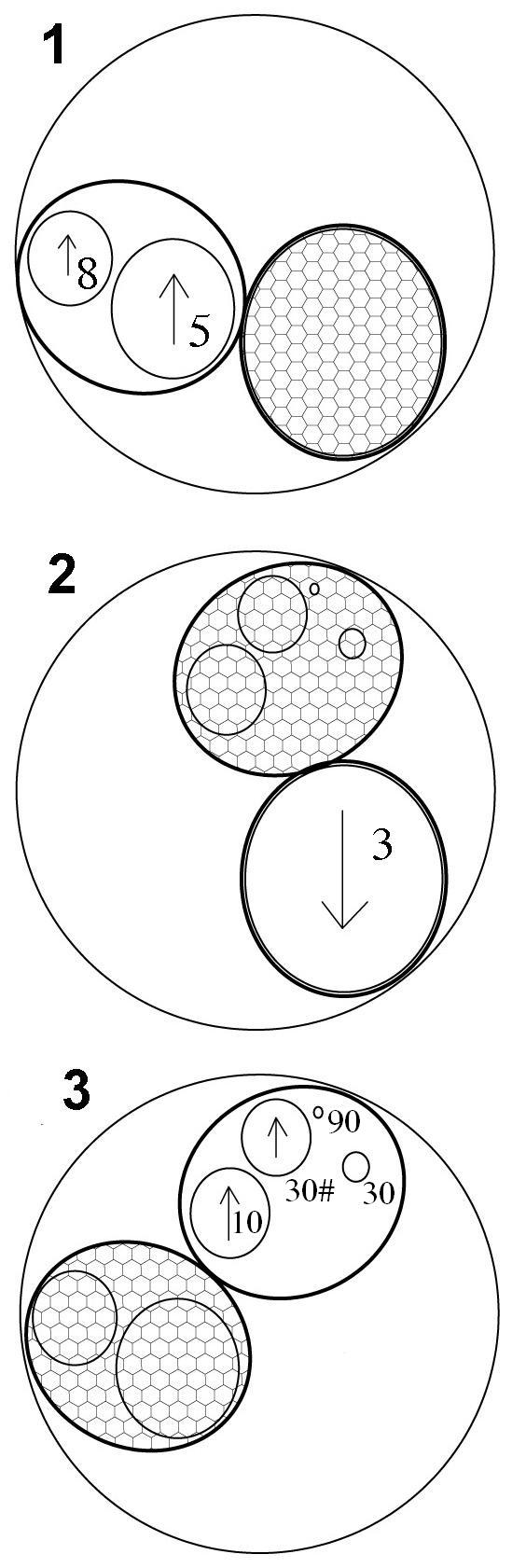}
 \caption{Layout of horn apertures on the aperture plate and the position of the sky port and external calibrator in the three sky viewing positions of the carousel.  The horns are arrayed in three groups, with the 3 GHz occupying one, the 5 and 8 GHz occupying another, and the high bands occupying the third.  The horns and are sliced at the apertures so that the beams point 30\degr\ from zenith in the directions indicated, with all horns pointing in the same direction with the exception of the 3 GHz, which points opposite.  For each of the carousel positions, the open ellipse shows the position of the sky port, and the filled ellipse shows the position of the external calibrator.  The arrows indicate the direction in which the horn beams point 30\degr\ from vertical.}
 \label{apfigure}
 \end{figure}

The corrugated horn antennas are arrayed on the aperture plate in three clusters, with the 3 GHz horn occupying one, the 5 and 8 GHz horns occupying a second, and the remaining horns, the ''high bands'', occupying the third.  All horns except the 3 GHz point 30\degr\ from zenith in one direction, while the 3 GHz points 30\degr\ from zenith in the opposite direction.  The sky port in the carousel is surrounded by reflective stainless steel flares which shield the edge of the antenna beams from instrument contamination and direct boiloff helium gas out of the port to discourage nitrogen condensation in the horn aperture.  Figure \ref{carousel} shows a photograph of the carousel being lowered onto the aperture plate, with the sky port and the radiometric side of the external calibrator visible.  Figure \ref{apfigure} shows the arrangement of the horn apertures.

The carousel is supported 1.5 mm above the aperture plate by wheel bearings on the edge and a Kynar$^\copyright$ plastic bearing in the center.  It is turned with a motor and chain drive, with the motor mounted outside of the dewar.  The motor is commanded to run and stops, with the engagement of a logic switch, when the proper alignment of the carousel relative to the aperture plate is reached, thus ensuring accurate and repeatable positioning of the carousel.  There are three carousel stopping positions for sky and external calibrator viewing, which are, in counterclockwise order, 1) the ''5 and 8 sky'' position, where the 5 and 8 GHz horns view the sky and the 3 GHz horn views the external calibrator, 2) the ''3 sky'' position in which the 3 GHz horn views the sky and the high band horns view the calibrator, and 3) the ''high sky'' position where the high band horns view the sky and the 5 and 8 GHz horns view the external calibrator.  The remaining group of horns viewing neither the sky nor the external calibrator in any of these positions are viewing a flat metal plate of the carousel.  A fourth stopping position is used on ascent to align a vent hole on the carousel with one on the aperture plate, thereby channeling boil-off gas to cool the back of the external calibrator.

Carousel alignment at each of the four positions is repeatable to within one millimeter on the outside circumference.  The sky port and external calibrator each have nearly the same area and shape as the aperture of the largest horn aperture (3 GHz).  Mis-alignment of more than 1 cm would allow a section of the aperture plate and flares to extend over the antenna aperture.  The actual alignment precision of 1 mm prevents such mis-alignment. 

\subsection{Gondola configuration}

\begin{figure}
\includegraphics[width=3.5in]{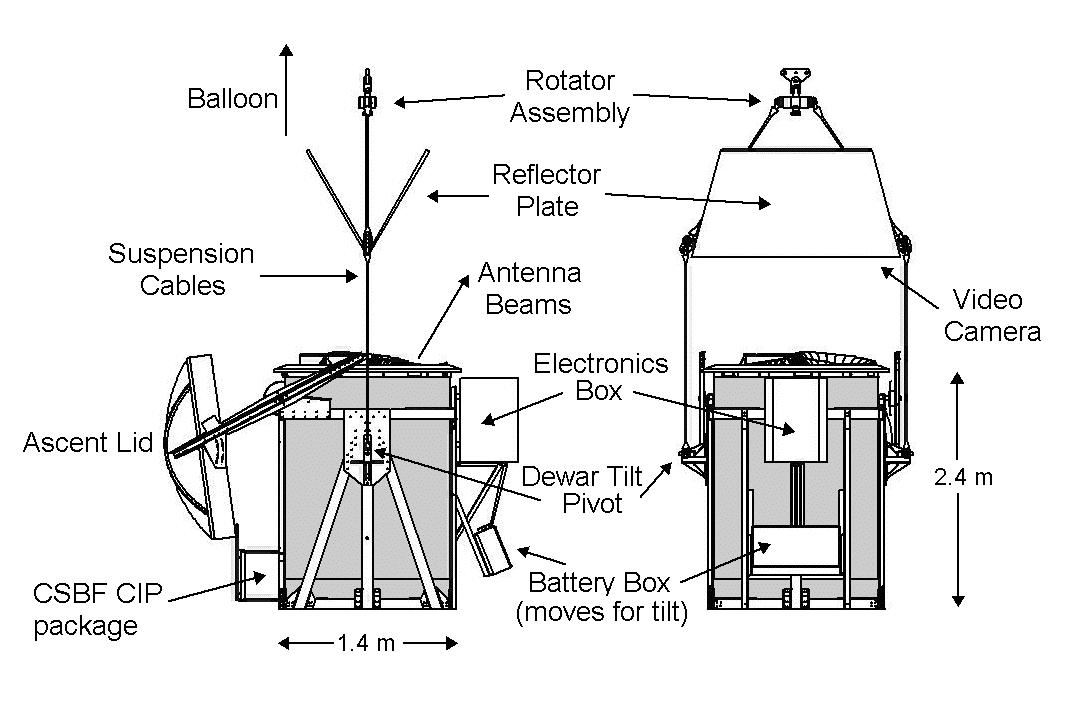}
\caption{ARCADE 2 gondola in 2006 flight configuration.  A deployable lid protects the cold optics during launch and ascent.  The reflective shield screening the flight train and parachute from view of the antennas, and the bar on which it sits, are the largest single sources of systematic uncertainty and is measured in flight by heating the shield.}
\label{gondola}
\end{figure}

\begin{figure}
 \includegraphics[width=2.5in]{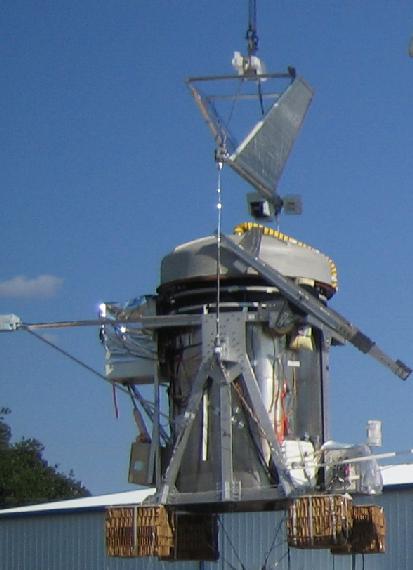}
 \caption{Photograph of the ARCADE 2 instrument just prior to 2006 flight.  The instrument core is contained within the large (1.5 m diameter, 2.4 m tall) bucket dewar.  The lid is shown closed for launch.  The electronics box containing the warm stages of the radiometers and the payload electronics is mounted to the left of the dewar, and the three-axis magnetometers to determine orientation relative to the Earth's magnetic field are seen to the left of the electronics box at the edge of the frame.  The reflector plate is visible at the top of the photo above the lid.  Cardboard pads at the four bottom corners absorb some of the impact when the payload hits the ground upon termination of the flight.}
 \label{payload}
 \end{figure}

Figure \ref{gondola} shows a schematic of the entire payload.  The dewar is mounted in an external frame supported 64 m below the balloon, and boxes containing the read out and control electronics and batteries are mounted on the frame.  The external frame is suspended by two vertical cables from a horizontal spreader bar 1.14 m above the top of the dewar, which itself is suspended by two cables from a rotator assembly.  The rotator maintains the rotation of the payload below the balloon at approximately 0.6 RPM.  The rotator assembly is suspended from a truck plate, above which is the flight train.  Reflector plates of metalized foam are mounted on the spreader bar to shield the edge of the antenna beam from the flight train.  Figure \ref{payload} shows a photograph of the payload prior to a launch.  The total mass at launch, including liquid helium, is 2400 kg.

The dewar can be tipped to angles of 2\degr\ from vertical, changing the angle of the antennas with respect to the reflector plates and flight train, by moving the battery box outward from the frame.  A fiberglass lid mounted on the frame is closed to cover the dewar on ascent and descent and opened for observations.  Thermometry, heater, and other signals are interfaced between the dewar and the exterior electronics box via cabling and a collar of insulated connectors at the top of the dewar.

Three-axis magnetometers and clinometers mounted on the frame, along with GPS latitude, longitude, and altitude data recorded by CSBF instruments, allow the reconstruction of the pointing of the antenna beams during flight.  During the 2006 flight, the magnetometers failed, and the pointing was reconstructed with a combination of the clinometers and radiometric observations of the Galactic plane crossings.  The uncertainty in the reconstructed pointing (a small fraction of a degree) is small compared to the beam size.
  
\subsection{\label{hornsection}Horn Antennas}
  
The corrugated antennas for six of the channels have 11.6\degr\ full width at half power Gaussian beams, while the antenna for the 30 GHz narrow beam channel has a 4\degr\ full width at half power Gaussian beam.  The horns are sliced at the aperture to point 30\degr\ from zenith when hung from the flat aluminum aperture plate.  This is so the antenna beam boresights are directed away from the flight train and so that they trace out a circle 60\degr\ on the sky as the dewar rotates below the balloon.  In order to achieve the narrowest possible beam at 3 GHz given the spatial constraints, a curved profiling of the horns was employed.  A full discussion of the horns is given by \citep{SH06}.

 \begin{figure}
\includegraphics[width=3.5in]{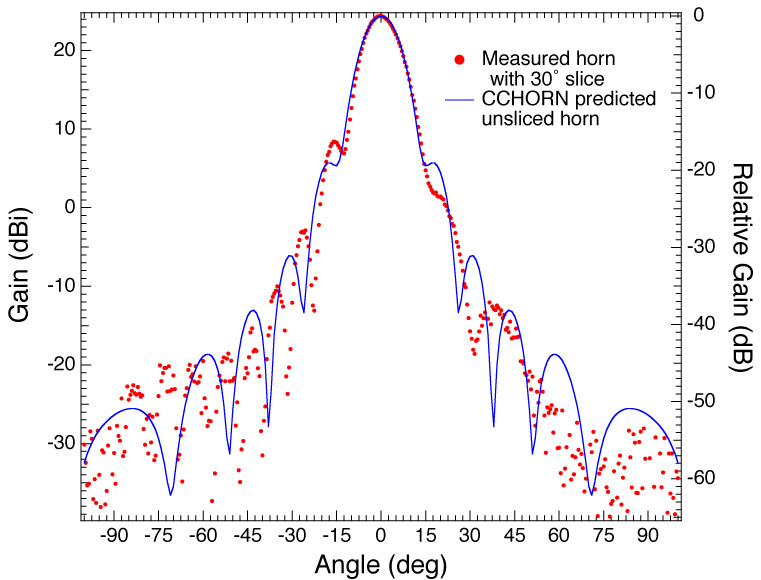}
 \caption{Measured response for the 10 GHz horn, at the center band frequency of 10.11 GHz (dots) as a function of angle, scanning in the plane where the effect of the 30\degr\ aperture slice is maximal.  The predicted simulated beam pattern for an unsliced horn is also shown (solid line).  The sidelobe response is low, and the effect of the slice is seen primarily in the first sidelobe.  Far sidelobes are further supressed.  The beam pattern does not vary appreciably over the frequency band. }
 \label{hornfig}
 \end{figure}

The horns were designed using mode matching simulation software.  The beam pattern for the fabricated 10 GHz horn was mapped over greater than 2$\pi$ steradian at the Goddard Electromagnetic Anechoic Chamber test range at the NASA Goddard Space Flight Center.  The measured beam pattern in the plane containing the maximal effect of the 30\degr\ aperture slice is shown along with the predicted one from design in Figure \ref{hornfig}.  The aperture slice has a minimal effect on the symmetry of the beam, and is seen primarily in the first sidelobe response, depressing the response on the long side and increasing the response on the short side.  This effect is a few dB, overlayed on a first sidelobe response that is 20 dB below the main beam.  Response in far sidelobes below the aperture plane is surpressed by more than 50 dB.  Because of the large physical sizes involved, it is not practical to measure the beam pattern for the lower frequency horns, or any of the horns once installed in the instrument, but the correctness of the simulated beam patterns, the low sidelobe response, and the negligable effect of the slice has been demonstrated with the 10 GHz horn beam map.  Furthermore, in the ARCADE I instrument, we were able to measure the beam of its 10 GHz horn both prior to and after installation in the instrument, and the far sidelobe response was identical \citep{KI04}, further demonstrating consistency.

\subsection{\label{radsection}Radiometers}

\begin{figure}
\includegraphics[width=3.5in]{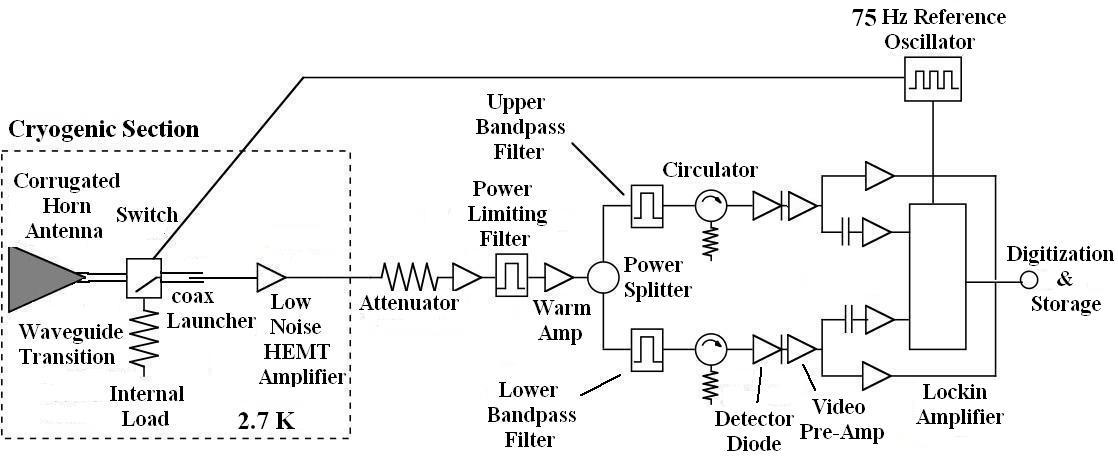}
\caption{Block diagram of ARCADE 2 radiometers.  The demodulated output is proportional to the difference in temperature between radiation from the antenna and the internal reference load.  The warm stage of the radiometers operate at $\sim$280 K. }
\label{radblock}
\end{figure}

Figure \ref{radblock} shows a block diagram of the radiometers.  Radiation incident from the horn goes through a compact circular to rectangular waveguide transition \citep{WT}.  Typical reflections from the horn and transition system are less than -30 dB across the entire radiometer band.  A switch chops at 75 Hz between radiation from the horn and that from the internal reference load.  At 3 and 5 GHz, a micro-electrical-mechanical system (MEMS) switch is used, while at the other channels a latching ferrite waveguide switch is used.  At 3 and 5 GHz, the internal reference load is a simple coax termination stood off with a stainless steel coax section.  At 8 and 10 GHZ, the reference load is a wedge termination in waveguide with a layer of Steelcast, a microwave absorber consisting of stainless steel powder mixed into a commercially available epoxy \citep{WS} cast onto an aluminum substrate.   At 30 and 90 GHz, the reference load is a split block wedge configuration of Steelcast in waveguide.  These wedge termination and split-block reference loads have reflected power attenuation of more than 30 dB across the entire frequency band and are described by \citet{WL}.  We measure the temperature of the absorber in each reference load with ruthenium-oxide resistance thermometers (see \S \ref{TTC}).

Each radiometer uses a cryogenic high electron mobility transistor (HEMT) based front-end amplifier which sets the system noise temperature.  The radiation exiting the switch is amplified by the cold HEMT and propagates, via coaxial cable at the low frequencies and waveguide at 30 and 90 GHz, out of the dewar to the warm stage contained in the electronics box.  The warm stage features a warm HEMT amplifier, an attenuator to eliminate reflections and tune the output power to match downstream components, a band pass filter to select the desired frequency band, a second warm HEMT amplifier, and a power divider to split the signal into a high and a low frequency channel.  Each of the high and low channels has a bandpass filter, a detector diode, and an audio frequency preamp, outputting a voltage level corresponding to the power of the radiation incident on the diode.  The voltage signal is then carried to an electronics board where a lockin amplifier demodulates the signal in phase with the switch, integrates it for .533 seconds, and digitizes it.  In this way, the final output is proportional to the difference in temperature between what the radiometer is viewing and the internal reference load, averaged over the integration period.  There is also a total power output signal in the data stream which is not demodulated by the lockin amplifier. 

 \begin{figure}
\includegraphics[width=2.5in]{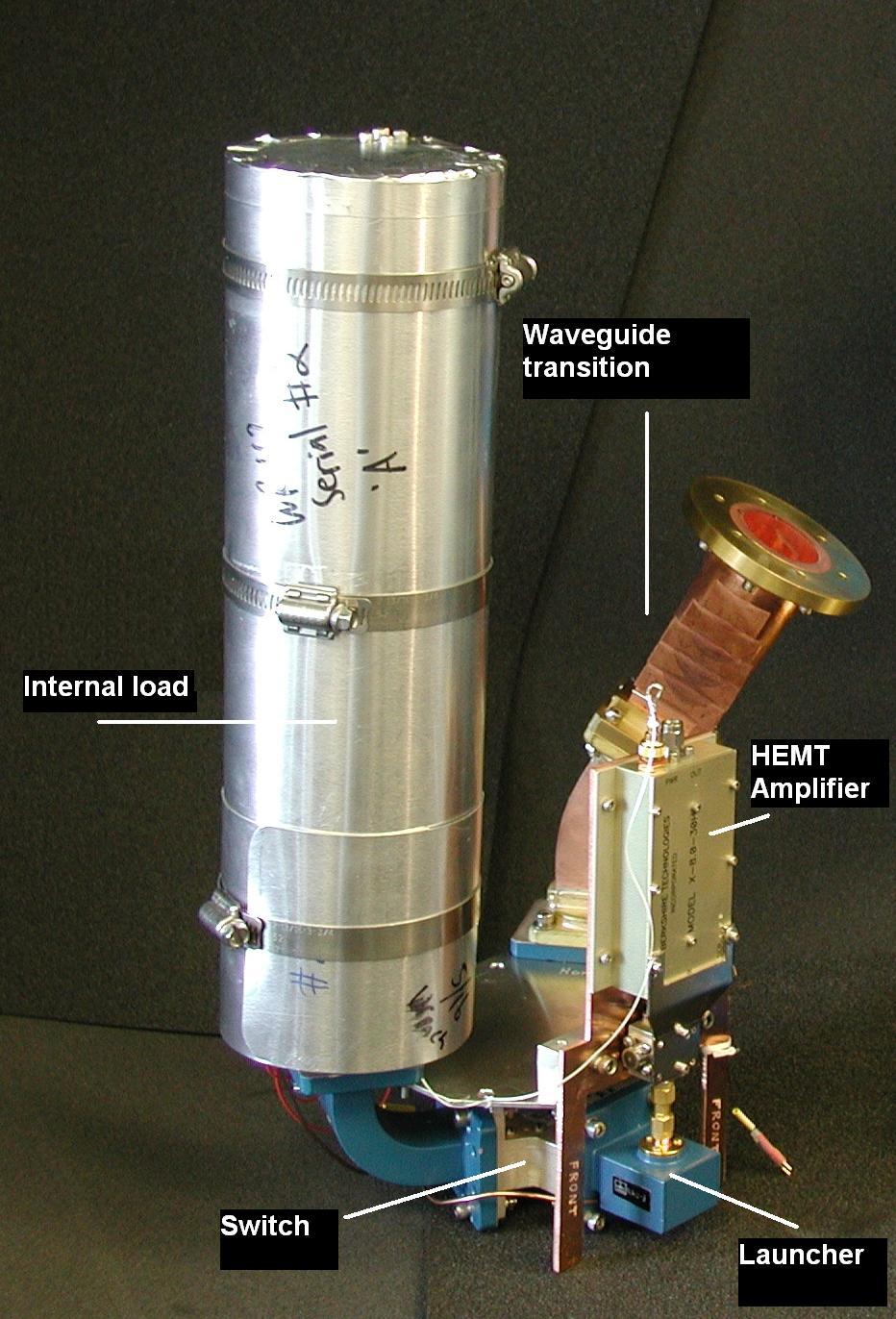}
 \caption{Photograph of cold stage of ARCADE 2 8 GHz radiometer.  The throat of the horn antenna bolts to the open circular waveguide end.  The internal reference load of this radiometer is a wedge termination in waveguide, which is surrounded by the insulative cylinder visible in the photograph.}
 \label{radcold}
 \end{figure}

Tables~\ref{tbl-1} and ~\ref{tbl-2} show selected radiometer properties and figures of merit.  Figure \ref{radcold} shows a photograph of a radiometer cold stage.  The custom designed radiometer components are the horns, the circular to rectangular waveguide transitions \citep{WT}, the ferrite waveguide switches, and the cold amplifiers at 3, 5, 8, and 10 GHz.  

The cold stages of the radiometers are housed in pans that can hold liquid helium, with a cascading pond system linking the pans.  The components are stood off from but thermally linked to the liquid, with the liquid providing cooling to maintain the components at temperatures below 4 K.  For the horn throats and internal loads, temperatures within this range are selected by SPID control, as described in \S \ref{TTC}, while the switches and HEMT amplifiers run near the liquid temperature.  

\begin{table*}[t]
\begin{center}
\caption{Table of selected radiometer hardware specifications\label{tbl-1}}
\begin{tabular}{crrrrrrrrrrr}
\tableline\tableline
 & 3 GHz & 5 GHz & 8 GHz & 10 GHz & 30 GHz & 30\#\tablenotemark{a} GHz & 90 GHz\\
\tableline
Low Band (GHz) &3.09-3.30 &5.16-5.50 &7.80-8.15 &9.2-10.15 &28.5-30.5 &28.5-30.5 &87.5-89.0\\
High Band (GHz) &3.30-3.52 &5.50-5.83 &8.15-8.50 &10.15-10.83 &30.5-31.5 &30.5-31.5 &89.0-90.5\\
Switch Type\tablenotemark{b} &MEMS &MEMS &Cir &Cir &Cir &Cir &Cir\\
Cold Amplifier Mfg. &Berkshire &Berkshire &Berkshire &Berkshire &Spacek &Spacek &JPL\\
Cold Amp Model \# &S-3.5-30H &C-5.0-25H &X8.0-30H &X10.0-30H &26-3WC\tablenotemark{c} &26-3WC\tablenotemark{d} &90 GHz Amps\\
Cold Amp Serial \# &105 &106 &108 &101 &5D12 &6C21 &W82$\&$W105\tablenotemark{d}\\
Cold Amp Gain (dB) &40 &29 &40 &40 &26 &20 &52\\
Warm Amp 1 Gain (dB) &35 &33 &33 &30 &46 &22 &-7\tablenotemark{e}\\
Attenuator (dB) &-26 &-13 &-26 &-20 &-3 &-10 &-6\\
Warm Amp 2 Gain (dB) &35 &33 &33 &31 &- &23 &40\\
Detector Power (dBm) &-21.3 &-20.6 &-28 &-24.6 &-25.9 &-42.5 &-24.6\\
X-Cal reflection\tablenotemark{f} (dB) & 42.4 & 55.5 & 58.6 & 62.7 & 55.6 & xx\tablenotemark{g} & 56.6\\
\tableline
\end{tabular}
\end{center}
$^a$ 30\# designates the 30 GHz channel with the narrower antenna beam.\\
$^b$ Switches are either MEMS (Micro-Electro-Mechanical System) or ferrite latching waveguide circulator switches (Cir).\\ 
$^c$ These model numbers have the prefix "SL315-".\\
$^d$ At 90 GHz there two are cold amplifiers in series.\\
$^e$ The 90 GHz channel warm stage uses a mixer, with a 79.5 GHz local oscillator, to translate it to 8-11 GHz, and all following components operate in that frequency range.\\ 
$^f$ This is the measured attenuation of reflections from the external calibrator, when it is viewed with the horn antenna for the channel.  This values presented here were measured in ground testing, and the measurement is described by \citet{F06} .\\
$^g$This external calibrator reflections were not measured directly with the 30\# channel antenna, but they are assumed to be even lower than for the other 30 GHz antenna.\\
\end{table*}

\subsection{External calibrator}

ARCADE 2 determines the radiometric temperature of the sky by using a full-aperture blackbody external calibrator as an absolute temperature reference.  To function as a blackbody emitter, it should have very low power reflection across the entire ARCADE 2 frequency range.  The emitting surfaces of the external calibrator need to be close to isothermal, and, while located on the carousel at the top of an open bucket dewar, have to be precisely temperature controlled within the range of 2.5 K to 3.1 K.   

The external calibrator is based on the successful model from COBE/FIRAS of a full aperture calibrator that is absorptive and isothermal, achieved through a combination of material and geometry \citep{M99}.  The ARCADE 2 calibrator is as good as the COBE/FIRAS one radiometrically, having a similar level of reflections, although the ARCADE 2 frequency coverage extends to much longer wavelengths.  The calibrator has a reflected power attenuation of less than -55 dB in the range from 5 to 90 GHz and of -42 dB at 3 GHz.  Measured reflected power attenuation values are presented in ~\ref{tbl-1} and the external calibrator's radiometric performance is described in detail by \citet{F06}.  

\begin{figure}
\includegraphics[width=2.5in]{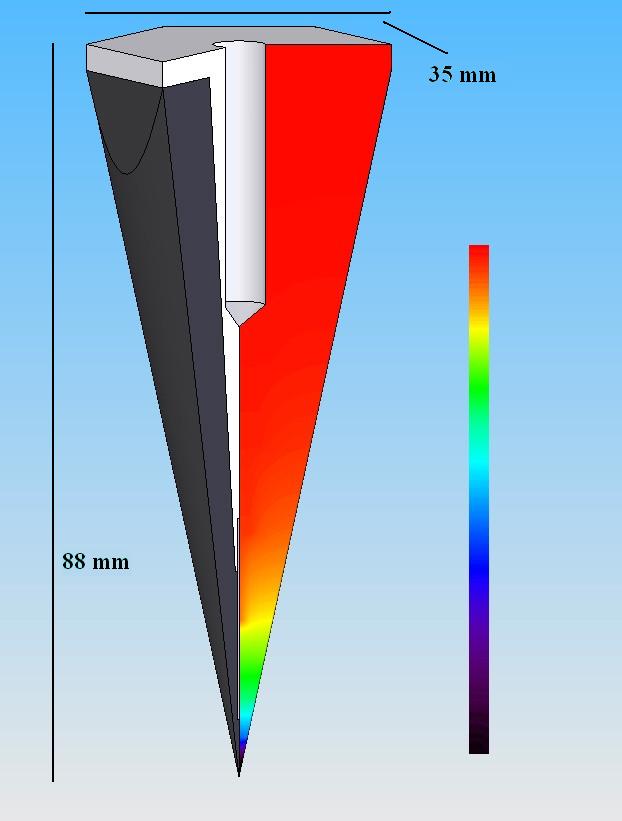}
 \caption{Cut away view of external calibrator cone showing internal structure and predicted linear temperature profile from the 2006 flight, with the base thermally fixed and the cone immersed in a cold atmosphere.  The cone consists of Steelcast absorber (black) cast onto an aluminum core (light grey).  There is also a copper wire running from the tip of the aluminum core to nearly the tip of the cone (darker grey).  The hole in the base is threaded for a mounting bolt to affix the cone to the aluminum back plate.  The measured thermal gradient is 600 mK from the base to the tip of the cones.  98\% of the total gradient is in a region near the tip containing 3\% of the absorber, and the mean depth for absorption varies with frequency.}
 \label{cone}
 \end{figure}

The calibrator consists of 298 cones, each 88 mm long and 35 mm in diameter at the base, of Steelcast absorber (see \S \ref{radsection}) cast onto an aluminum core.  The aluminum provides for enhanced thermal conductivity and therefore reduced thermal gradients, which is augmented by a copper wire epoxied onto the end of the aluminum core and running almost to the tip.  The radiometric side of the external calibrator is visible in Figure \ref{carousel}.  Figure \ref{cone} shows a cut away view of a cone.  

The cones are mounted pointing downward on a horizontal aluminum plate, behind which are 50 alternating layers of alloy 1100 aluminum and fiberglass sheets, to give a low vertical but high horizontal thermal conductivity, to the end of getting the different cones as isothermal as possible.  This aluminum plate is welded to an aluminum shielding which surrounds the elliptical circumference of the area of cones so that the cones are surrounded on the top and sides by an isothermal metal surface maintained at temperatures near 2.7 K.  Behind the stack of alternating layers is another horizontal aluminum back plate, and behind that a half inch gap to allow room for bolt heads, heating elements, and wiring.  In the 2006 configuration the entire calibrator is surrounded, except on the radiometric side, by a tank of liquid helium, which is thermally coupled with stainless steel standoffs to the rear aluminum plate.  This layer of liquid helium intercepts any external heat load incident on the external calibrator from the back and sides. 

Temperatures are monitored in 23 of the cones with thermometers embedded in the Steelcast absorber at varying depths and radii.  Three of the cones contain two thermometers each, for a total of 26 thermometers within the absorber.  Resistance heaters are mounted on the rear aluminum plate, and 7 additional thermometers are affixed at various points.  Thermometry and heater signals from the calibrator, as well as from elsewhere on the carousel, are carried through a hollow tube in the axis of the carousel to the rest of the core inside the dewar.  The true emission temperature of the external calibrator is a volume integral over the physical temperature distribution weighted by the absorber emissivity and the electric field distribution from the antenna aperture.  We approximate the true emission temperature as a linear combination of the temperatures as measured by the thermometers, with weights derived from the actual flight data as described by \citet{F08}.

\begin{table*}[tbph]
\begin{center}\caption{Table of selected measured radiometer performance specifications from the 2006 flight\label{tbl-2}}
\begin{tabular}{rrrrrrrr}
\tablewidth{7.5in}
& Cntr Freq & Bandwidth & $T_{rcvr}$\tablenotemark{a} & Offset\tablenotemark{b} & Noise (pre-flt)\tablenotemark{c} & Noise (rsduls)\tablenotemark{d} & Noise (map)\tablenotemark{f} \\
Channel & (GHz) & (MHz) & (K) & (mK) & (mK$\sqrt{s})$ & (mK$\sqrt{s})$ & (mK$\sqrt{s})$ \\ \hline
3 GHz Lo & 3.15 & 210 & 5.5 & 180 & 7.1 & 9.3 & 11.8 \\
3 GHz Hi & 3.41 & 220 & 6.5 & 35 & 7.1 & 7.8 & 10.1 \\
5 GHz Lo & 5.33 & 340 & 6 & -210 & 3.2 & - & - \\
5 GHz Hi & 5.67 & 330 & 6 & -200 & 3.5 & - & - \\
8 GHz Lo & 7.97 & 350 & 10 & 6 & 1.4 & 5.5 & 5.5 \\
8 GHz Hi & 8.33 & 350 & 8 & 11 & 1.4 & 6.1 & 5.2 \\
10 GHz Lo & 9.72 & 860 & 13 & 180 & 6.8 & 3.7 & 3.0 \\
10 GHz Hi & 10.49 & 680 & 11 & 35 & 5.3 & 3.7 & 3.0 \\
30 GHz Lo & 29.5 & 2000 & 75 & -30 & 21.5 & 208.2 & 206.9 \\
30 GHz Hi & 31.0 & 1000 & 72 & -15 & 14.9 & 103.3 & 103.4 \\
30\# GHz Lo & 29.5 & 2000 & 270 & 32 & 18 & 885.0 & 880.4 \\
30\# GHz Hi & 31.0 & 1000 & 340 & 38 & 14 & 418.5 & 406.1 \\
90 GHz Lo & 88.2 & 1500 & 44 & -75 & 5.2 & 50.5 & 42.0 \\
90 GHz Hi & 89.8 & 1500 & 38 & -95 & 5.7 & 25.3 & 27.1 \\
\hline
\end{tabular}
\end{center}
$^a$ T$_{rcvr}$ is the receiver noise temperature of the amplifier, a figure of merit that is equal to the temperature that would be observed by a total power radiometer containing the amplifier and viewing a source at a temperature of absolute zero.  This values presented here were measured in ground testing prior to the 2006 flight.\\
$^b$ The constant offset is the radiometer output when the internal reference load and the object being viewed are at the same temperature, multiplied here by the gain to be expressed as a temperature.\\
$^c$ This is the white noise as measured in ground testing prior to the 2006 flight.\\
$^d$ This is the white noise determined from the residuals of the data from the 2006 flight.\\ 
$^e$ This is the white noise as determined from the sky map variance in the 2006 flight.\\
\end{table*}

\subsection{Payload electronics}

The main payload electronics consists of the cryogenic thermometer resistance readout and control for heaters, as described in \S \ref{TTC}, as well as radiometer lockin and integration, voltage readout for the various payload analog devices, including the magnetometers, clinometers, and ambient temperature transducers, digital logic level readout, generation of the switch driving current, and generation and amplification for commandable signals to the lid, rotator, tilt, and carousel movement motors.  These functions are performed by custom electronics boards.  Typical power required for the electronics is 220 W, with peak capacity 1800 W.

The digital data stream is relayed via the RS-232 serial data standard to the Consolidated Instrument Package (CIP) provided by CSBF, which transmits it, along with data from the CSBF instruments and the video signal, to the ground.  Each 1.067 s record of data consists of digitized counts corresponding to a voltage reading across every cryogenic thermometer, the voltage reading across the reference resistors on each thermometer readout board (see \S \ref{TTC}), the voltage output of the various analog payload devices, the SPID control parameters for every controlled heater, digital logic levels, and two readings of the demodulated and total voltage output of every radiometer channel.  Each 1.067 s also allows four two byte commands to be transmitted via the CIP to the instrument and executed.  The commands include the setting of SPID parameters, lockin amplifier gains, and motor movement.

A video camera mounted on the spreader bar above the dewar allows direct imaging of the cold optics in flight.  Two banks of light-emitting diodes provide the necessary illumination.  The camera and lights can be commanded on and off, and we do not use data for science analysis from times when they are on.

\subsection{Flight operations}

When the instrument is ready for flight, the lid is closed and the dewar is cooled with nitrogen to around 100 K.  We then fill the dewar with around 1900 liters of liquid helium, which takes several hours.  We await a launch opportunity, with the helium level topped off each day.  Approximately 200 litres of liquid helium are boiled off per day. To avoid freezing the lid to the dewar with ice accumulation while the payload awaits flight, we position an outflow hose leading from a vent hole in the top of the lid to a position close to level with the dewar rim.  This maintains some positive pressure and helium gas outflow at the lid-dewar interface, which discourages the condensation of ambient water vapor.

We launch the instrument with the carousel in the ''ascent'' position that aligns the vent holes in the aperture and carousel, which directs boil-off gas across the back of the external calibrator, providing a powerful cooling source for its large thermal mass.  On ascent, about 1/3 of the helium is boiled off, with the remainder cooled continuously to 1.5 K when float altitude is reached.  We turn on the superfluid pumps once the helium bath is below the superfluid transition temperature of 2.177 K.  After three hours of ascent, float altitude is reached and we open the instrument lid for observing.  During observing, we move the carousel to a new position about once every five minutes.  We observe for around four hours, limited by the westward drift of the balloon out of range of telemetry.  The lid is closed for descent, and the payload is severed from the balloon and returns to the ground on a parachute, to be recovered by CSBF staff.  At the end of the 2006 flight, nearly 800 liters of liquid helium remained in the dewar, or enough for 6 hours more of observation.

In the 2005 flight, the carousel became stuck in one position soon after observing commenced.  The cause was traced to the output torque of the carousel motor exceeding the maximum torque of the attached gear box, stripping the gears, and was remedied for the 2006 flight.  In the 2006 flight, the most significant instrumental failure was with the 5 GHz MEMS switch, rendering data from that channel not useful for science analysis.

\section{Cryogenic Performance}

The ability to measure and control the temperature of cryogenic components while in the presence of a variable helium gas flow and potentially exposed to ambient sources of warming is the limiting factor in the precision of radiometric temperature measurements made with ARCADE 2.  Thermal measurement and control is the most important and challenging facet of the experiment.

ARCADE 2 requires component temperatures to be maintained near 2.7 K.  Some components are controlled passively by being thermally sunk to liquid helium and exposed to cold helium gas.  Other components are actively temperature controlled, with both a coupling to liquid helium and controllable resistance heaters.  The gas cooling is an undesirable perturbative effect on actively controlled components.  Liquid helium is moved to needed areas outside of the liquid helium bath, such as the aperture plate and carousel at the top of the dewar and the cold stages of the radiometers.  At ballooning altitudes liquid helium is well below the superfluid transition temperature and does not respond to mechanical pumping, so superfluid pumps with no moving parts are used.  In such pumps, with heating power of less than 1 W, liquid can be pumped to a height of several meters.  ARCADE 2 features 13 superfluid pumps which can move 55 liters per minute to a height of 2.5 m.

\subsection{Thermometry and temperature control} \label{TTC}

We measure 104 cryogenic temperatures in the ARCADE 2 payload.  26 are within the external calibrator cones, 7 are elsewhere on the calibrator, another 19 are at various points on the carousel, 26 are on components of the radiometers including the internal reference loads, switches, horns, and the pans in which the radiometers sit, and 26 are at various other points within the dewar.  We measure these temperatures using four-wire AC resistance measurements of ruthenium-oxide resistors, whose resistance is a strong function of temperature below 4 K.  The thermometers are excited by a 1 $\mu$A 37.5 Hz square wave current, and signals are carried through cold stages via a sequence of brass, copper, and manganin cryo-wire.  The resistances of the thermometers are read by custom electronics boards \citep{F02}, which output a digitized voltage level for every thermometer in every data record.  The boards also contain five on-board resistors of known resistance spanning the dynamic range of the cryogenic thermometer resistances that are excited with the same current as the cryogenic thermometers.  In this way, a reliable digital voltage counts to resistance conversion is obtained in every data record to express the measured resistance of the cryogenic thermometers.  The measured signals can then be converted in software to temperatures using look-up tables containing the resistance versus temperature curves for each thermometer. 

The resistance versus temperature curves for the ruthenium-oxide thermometers are determined in ground testing against a thermometer of identical design previously calibrated by NIST.  This style of thermometers has demonstrated calibration stability to within 1 mK over four years \citep{KI04}, verified with the observed lambda superfluid transition as an absolute reference.  Thermometer self-heating is negligible (less than 1 nW) and is included in the resistance-to-temperature calibration process.

\begin{figure}
\includegraphics[width=3.5in]{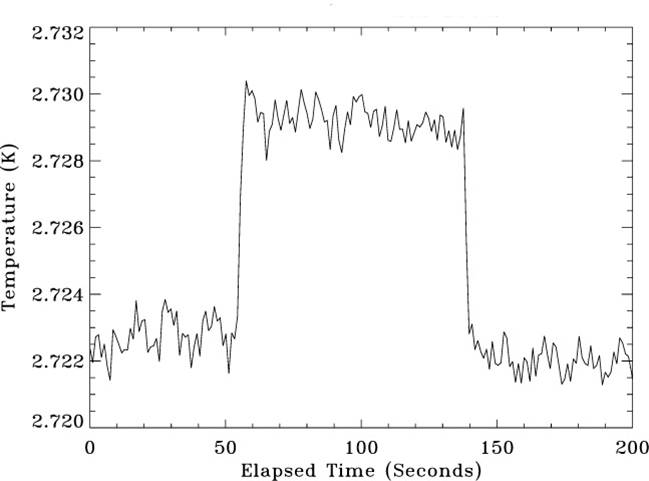}
\caption{Temperature of the 5 GHz internal reference load vs. time for a section of data from the 2006 flight showing SPID temperature control of the load.  The load has a relatively small mass and is located in a steady thermal environment.  The temperature was commanded to be set to 2.730 K at 51 seconds, and back to 2.724 K at 138 seconds.  In general, it is important that the temperature of components under SPID control be steady and known, but not that specific commanded temperatures be obtained exactly. }
 \label{5spid}
 \end{figure}

\begin{figure}
\includegraphics[width=3.5in]{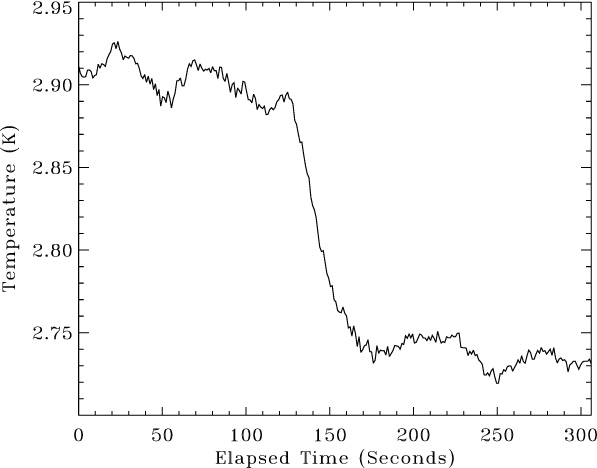}
 \caption{ Temperature vs. time of a cone in the external calibrator for a section of data from the 2006 flight showing SPID temperature control of the external calibrator.  The external calirator is a relatively large system exposed to changeable thermal conditions.  The ringing is beneficial as it is slow compared to the radiometer response time and allows the sampling of a wider range of calibrator temperatures. }
 \label{tgtspid}
 \end{figure}

Desired temperatures are maintained in key places, such as the radiometer internal reference loads and switches, horn throats, and external calibrator, through resistance heaters under SPID (set point, proportional, integral, and differential) control.  The values S,P,I, and D are set in real-time by the user for each of 32 SPID channels, and the desired set point temperature as well as the current actual temperature are expressed in counts, with the user performing the conversion between counts and temperature as necessary.  The output voltage level for each SPID channel is recalculated once per record (1.067 s) in firmware.  Figure \ref{5spid} shows the accuracy and precision of the SPID temperature control of the 5 GHz internal reference load.  This internal reference load has a relatively small mass (\verb1~110 g) and is located in a relatively stable thermal environment.  Figure \ref{tgtspid} shows the SPID temperature control of the external calibrator, a much larger system that is subject to changeable thermal dynamics.   

\subsection{External calibrator thermometer calibration}

Special care is taken in determining the resistance versus temperature curves for the ruthenium-oxide thermometers embedded in the cones of the external calibrator, as errors and uncertainties in these curves lead directly to errors and uncertainties in the measured radiometric temperature of the sky.  A specific calibration setup is employed for these thermometers to minimize any possible thermal gradients and wiring differences between calibration conditions and flight conditions.  In this configuration, the cones and a NIST calibrated standard thermometer are mounted on an 1100 series aluminum plate inside of an evacuated stainless steel pressure vessel, which is itself submerged within the liquid helium bath of a large (.3 m diameter, 1.5 m tall) test dewar.  The aluminum plate is thermally linked to the bath through a copper rod which passes through a superfluid-tight hole in the pressure vessel.  The liquid helium bath of the test dewar is pumped in stages, with the pumping valve opened some amount and then the bath left alone to eventually equilibrate to some steady temperature.  In this way, once a steady bath temperature is reached, with the evacuated vessel submerged in an isothermal liquid helium bath, there are no obvious sources of thermal gradients in the system.  The procedure leads to a steady thermal situation such that thermometer outputs show no coherent movement over timescales of ten minutes.  The aluminum plate, and therefore the cones and NIST standard thermometer, should be quite isothermal when the bath has equilibrated at a given temperature, as the aluminum plate is thermally lined to the bath and within an evacuated vessel with the walls at nearly the same temperature as the plate.

This process leads to discrete temperature points where a definite temperature, read by the NIST standard thermometer, can be associated with a measured resistance for each of the cone thermometers.  The fourteen total calibration points from three separate runs can be combined to form a reasonably dense calibration curve with values between 2.5 and 4.2 K.  

 \begin{figure}
\includegraphics[width=3.5in]{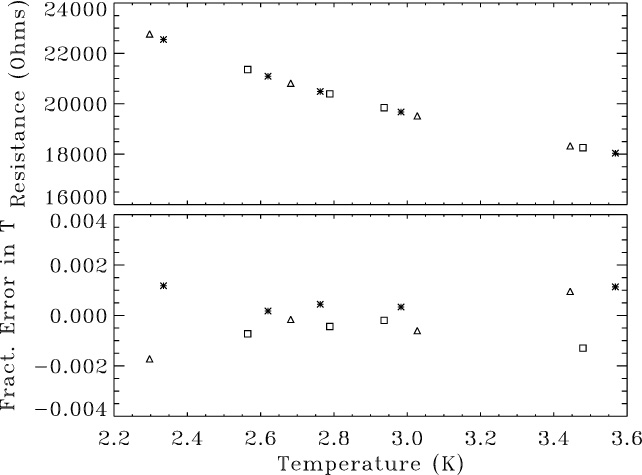}
 \caption{The upper plot shows a resistance versus temperature curve for a typical external calibrator cone thermometer.  The lower plot shows the fractional error in temperature from removing each temperature point in sequence to form a reduced information curve and feeding that reduced curve the resistance of the removed temperature point.  We take the magnitude of this error as an estimate of the uncertainty in the resistance versus temperature curve arising from sources other than statistical uncertainty and the capacitance correction.  The different shaped points represent data from the three different calibration runs.  }
 \label{caltherm}
 \end{figure}

 As we measure the resistance of the thermometers using a 37.5 Hz square wave excitation, this measured resistance includes the effect of shunt capacitance in the harnessing connecting the wires to the board.  Our thermometer calibration procedure assigns a temperature to each measured resistance using identical harnessing and electionics boards as in flight., so the effect of the shunt capacitance is automatically acounted for in the resistance vs. temperature curve obtained for each thermometer.  However, the NIST calibrated thermometer, which is read by the flight thermometer boards and flight harnessing in our calibration setup, was itself calibrated with DC readout techniques featuring no shunt capacitance.  To correct for the small effect on the measured resistance of the NIST calibrated thermometer, an estimate of the effect of the shunt capacitance of the calibration setup is formed using data taken with 20 resistors of known value placed at the end of the harnessing.  This yields a linear relationship between the actual resistance and the error in the measured resistance.  The shunt capacitance is around 300 pF which results in an offset in the measured resistance of 8 $\Omega$ when measuring a 20000 $\Omega$ resistor, which is the resistance of a thermometer at around 2.7 K.  An 8 $\Omega$ offset results in a 2 mK offset in the thermometer reading.  This small estimated effect is then subtracted from the raw NIST calibrated thermometer data, with the standard deviation of the linear fit providing the uncertainty in this applied correction.  

The total uncertainty in the determined resistance versus temperature curves of the cone thermometers consists of contributions from raw statistical uncertainty, from the uncertainty in the shunt capacitance correction for the NIST calibrated thermometer, and that from any other sources.  The first two are straightforward to determine, with statistical uncertainty well below 1 mK except at above 3.4 K where it is 1 mK, and the capacitance correction uncertainty well below 1 mK at temperatures below 3 K and rising to 3 mK at 3.4 K.  We estimate the remaining uncertainty from other sources using the figure of merit of how well data from the three separate test runs agree with each other.  Each run contains only four or five discrete temperature points.  We combine all 13 points to form a single resistance vs. temperature curve for each thermometer.  To estimate the remaining uncertainty, we remove each temperature point in sequence and compare the point to the curve determined from the 12 remaining points.  This is a test of the consistency of the combined curve and of the three runs.  Figure \ref{caltherm} shows a resistance versus temperature curve for a typical external calibrator cone thermometer, and the fractional error in temperature at each temperature point determined by removing the point.  

\begin{figure}
\includegraphics[width=3.5in]{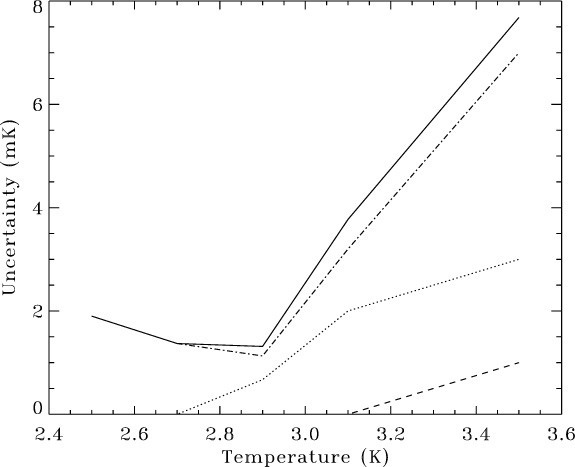}
 \caption{Uncertainty in the resistance versus temperature curve determination for the cones of the external calibrator, as a function of temperature.  The result presented is averaged over all of the cones.  The dashed line is statistical uncertainty, the dotted line is the uncertainty due to the applied correction to the NIST calibrated thermometer for shunt capacitance in the read-out wiring, and the dash-dot line is the magnitude of the error in reproducing the temperature of a calibration point given the resistance of that point and a reduced information curve not containing the point.  The solid line is the quadrature sum of the three sources, which is taken to be the total uncertainty.}
 \label{calthermerrors}
 \end{figure}

We take the magnitude of the error in reproducing the temperature of the removed point given the reduced information curve as an estimate of the uncertainty of that point in the curve, neglecting statistical and capacitance correction uncertainty.  Figure \ref{calthermerrors} shows an average over all cones of the uncertainty due to all three sources and the total uncertainty, taken to be the quadrature sum.  The uncertainties in temperature generally increase steeply above 3 K because at higher temperatures the resistance versus temperature curves flatten.  At 2.7 K, the average total uncertainty in the resistance versus temperature curves is 1.3 mK.

\subsection{Heat flow}

ARCADE 2 successfully maintains critical components near 2.7 K at and near the top of an open bucket dewar at 37 km altitude.  This is achieved by moving large quantities of superfluid liquid helium so that each component is either passively controlled at the bath temperature, or has a thermal path to the liquid to allow active temperature control.  Other components of the instrument are not thermally controlled, but generally come to temperatures between 1.5 K and 20 K depending on their location relative to external heat sources. We monitor temperatures on these non-critical components to demonstrate that their fluctuations do not affect the radiometer output.  The overall thermal performance of the entire instrument is well measured and is understood.

The thermal behavior of the instrument core in flight results from the balance of warming and cooling power.  Cooling is provided by both pumped liquid helium and boil-off helium gas.  The boil-off gas is channeled out of the dewar through the 38 mm perimeter gap between the aperture plate and the inner dewar wall.  There is also one 100 mm diameter gas vent port each on the aperture plate and carousel, which, when aligned, channel boil-off gas up across the back plate of the external calibrator and out through a path in the foam insulation covering the carousel.  This configuration is used to direct maximum helium gas flow to the calibrator back upon ascent.  

The sources of warming in flight are A) infrared radiation and conduction from the warmer ambient atmosphere, B) radiation and conduction from warm components of the instrument such as the dewar walls or the carousel rotation drive chain, and C) SPID heaters as described above.  There are three distinct paths by which ambient warmth can heat the core: 1) warming of the carousel from above, 2) warming of the aperture plate from above through the sky port of the carousel, and 3) warming via radiation from the warm wall of the dewar near the top.  To reduce the heat load from the dewar wall, there are three concentric sheet aluminum baffles inside the dewar wall extending .6 m down from the top of the dewar.  In addition, at the very top there is a five layer Vacuum Deposited Aluminum with Dacron concentric veil extending 200 mm down from the top outside of the outermost baffle.  To reduce the heat load to the carousel from above, it is covered in ~1 m$^3$ of foam insulation (Fomo Handi-Foam SR, a two-component slow-rise polyurethane foam), which is topped with a .25 mm reflective aluminum sheet.  

In the 2006 configuration, the aperture plate was cooled directly by six pump-fed liquid helium tanks mounted on the underside in various places.  Being relatively thin and covering a large area, the aperture plate is subject to thermal gradients even with liquid helium tanks in contact at discrete places.  The large horns at 3,5, and 8 GHz have sheet metal 'collars' on the exteriors, providing a sheath for liquid helium to maintain each horn aperture at the bath temperature.  Aperture plate temperatures in flight vary between 1.5 K and 10 K depending on location and time, while temperatures on the carousel vary between 4 K and 20 K.  In general, the aperture plate is colder than, and cools, the carousel, as the aperture plate is less warmed by heating from above and is cooled by helium tanks.

The major temporal temperature variations in flight result from differing gas flow dynamics in the different positions of the carousel.   In the position in which the 5 and 8 GHz horns are viewing the sky, the vent holes of the aperture plate and carousel are almost aligned, causing increased flow through the holes and therefore decreased flow to the perimeter and a resulting warming of perimeter areas of the aperture plate and carousel.  In the other two carousel positions this effect is not present and a sub-dominant effect is observed.  The "high sky" position exposes the 10 through 90 GHz horns to the sky.  Because these horns have relatively small apertures there is more metal of the aperture plate exposed to the sky port and therefore more infrared radiation incident on it from above.  This effect is not present when the 3 GHz horn views the sky, as the 3 GHz horn aperture completely fills the sky port and is maintained at the bath temperature by direct contact with the superfluid liquid helium collar on the exterior of the horn.  

There are also transient effects when the carousel moves from one position to another.  Portions of the carousel perimeter that pass near where the carousel drive chain enters through the connector collar are momentarily warmed, and the back plate of the external calibrator is momentarily cooled as the calibrator passes over the aperture plate vent hole, briefly channeling gas between the calibrator's aluminum shielding and the surrounding liquid helium tank and then across the back plate.  These effects on radiometrically active parts of the instrument used for science analysis are small, and data from times in which the carousel is moving are not used for science analysis.

The observed temperature gradient among the concentric baffles inside the dewar wall, from 9 K at the inner baffle to around 30 K at the outer one, indicates that the heat leak to the aperture plate from the warm dewar walls is reduced, via the baffling and blow-by of boil-off gas, to the level of a few mW.  The observed gradient through the foam topping the carousel indicates that the heat load from above to the carousel is on the order of 30 W, and this is roughly consistent with the source being infrared radiation from a 200 K body incident on the aluminum sheet topping the foam. 

In the 2006 flight, we experienced problems with temperature controlling the 3 and 8 GHz internal reference loads, both becoming fixed at near the liquid helium bath temperature for significant periods of the flight.  In the case of the 3 GHz load, this was likely caused by the insulative housing falling off, which was observed upon completion of the flight.  The 5 GHz load, which was of the same construction as the 3 GHz load, did not experience this problem, most likely because the 5 GHz load was located above the liquid Helium level during ascent.  

\subsection{External calibrator thermal performance}

In both the 2005 and 2006 flights, all radiometrically active parts of the external calibrator were maintained within 300 mK of 2.7 K.  However, the external calibrator cones displayed significant gradients from base to tip.  In the 2005 flight, the aperture plate, which was at temperatures significantly warmer than the calibrator, warmed the tips relative to the base.  The situation was reversed in the 2006 flight, where the presence of the underside liquid helium tanks caused the aperture plate to run significantly colder and the tips of the cones were cold relative to the bases.  The wrap-around tank of liquid helium surrounding the external calibrator successfully intercepts all heat loads from the top or sides.  The only un-controlled thermal link is between the cones on the radiometric side of the calibrator and the horn apertures and aperture plate, which is responsible for the thermal gradients observed in the calibrator.  In the 2006 flight, the maximum base to tip gradient was 600 mK.  Figure \ref{cone} shows a predicted temperature profile based on this gradient.  In light of the observed effects, a more uniform temperature could be achieved through active temperature control of the aperture plate with heaters to maintain it at near 2.7 K.

\subsection{Atmospheric condensation}

\begin{figure}
\includegraphics[width=3.0in]{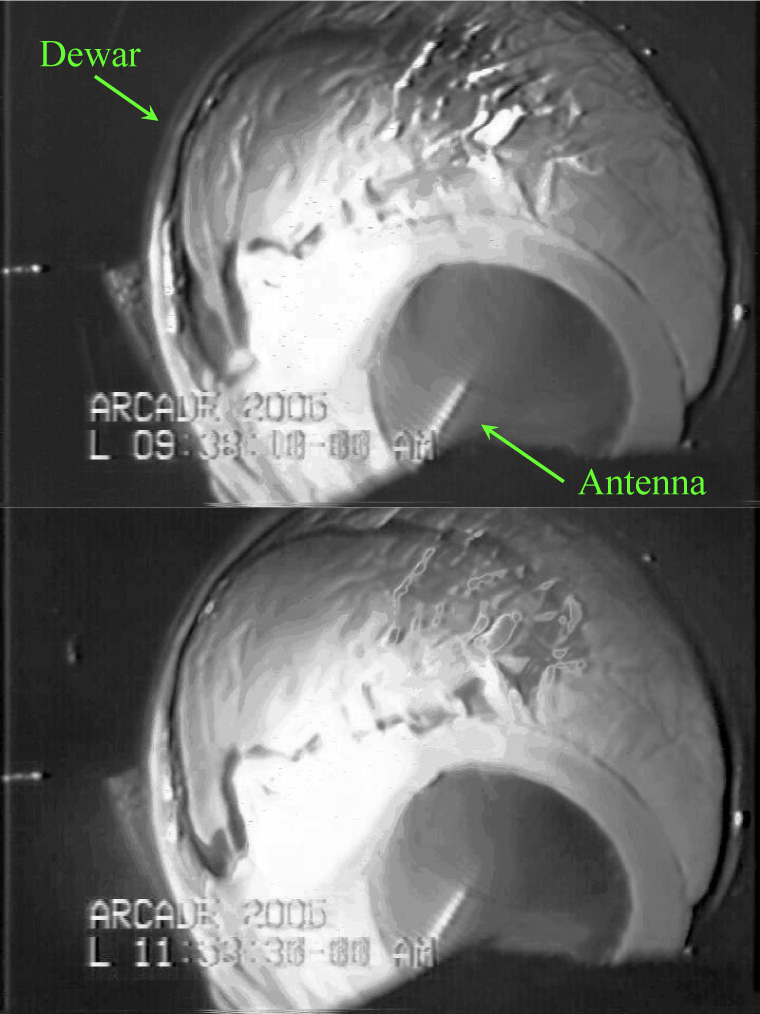}
\caption{Time averaged video camera images of the dewar aperture during two periods of the 2006 flight where the 3 GHz horn antenna was viewing the sky.  The time and date stamps in the video image do not correspond to the actual time and date.  The bottom image was taken two hours after the top image.  As can be seen by comparing the faintly visible grooves, there is no visible condensation in the horn antenna aperture over a two hour period.}
\label{aperture}
\end{figure}

The potential problem with a cold open aperture is condensation from the atmosphere.  Condensation on the optics will reflect microwave radiation adding to the radiometric temperature observed by the instrument in an unknown way.  In the course of an ARCADE 2 observing flight, the aperture plate and external calibrator are maintained at cryogenic temperatures and exposed open to the sky for over four hours.  Figure \ref{aperture} shows time averaged video camera images of the dewar aperture taken two hours apart during the 2006 flight.  No condensation is visible in the 3 GHz horn aperture despite the absence of any window between the horn and the atmosphere.  It is seen that the efflux of cold boiloff helium gas from the dewar is sufficient to reduce condensation in the horn aperture to below visibly detectable levels.  

\begin{table*}[tbph]
\begin{center}
\caption{Estimates of the radiometric temperature contribution from the balloon and other components visible in the antenna beams, for channels used for science analysis in the 2006 flight. See \S \ref{fltrain}.  All estimates are in mK.\label{tbl-3}}
\begin{tabular}{lrrrrrrrrrr}
\tablewidth{7.5in} Component&3L&3H&8L&8H&10L&10H&30L&30H&90L&90H \\
\hline
Balloon&0.1&0.1&0.2&0.2&0.3&0.3&1.9&2.0&14.8&14.8\\
Reflector shield&1.5&0.9&4.2&4.5&0.5&0.5&0.8&0.8&0.8&0.7\\
Lights\tablenotemark{a}&0.3&0.2&0.0&0.0&0.0&0.0&0.0&0.0&0.0&0.0\\
Spreader bar&8.5&4.3&31.4&36.6&1.7&1.1&1.3&1.7&1.3&1.1\\
Upper Suspension\tablenotemark{b}&0.1&0.0&0.2&0.2&0.0&0.0&0.0&0.0&0.0&0.0\\
Lower Suspension\tablenotemark{b}&0.0&0.0&0.2&0.2&0.0&0.0&0.0&0.0&0.0&0.0\\
Cold Flare\tablenotemark{c}&0.3&0.3&0.3&0.5&0.3&0.3&0.3&0.3&0.3&0.3\\ \hline
Total&10.8&5.8&36.6&42.2&2.9&2.3&4.4&4.8&17.2&16.9\\
\hline
\end{tabular}
\end{center}
$^a$ A bank of lights can be commanded on for when the video camera views the aperture.  Data when the lights are on are not used for science analysis.  This emission estimate is for the lights off.\\
$^b$ The lower suspension cables suspend the dewar from the spreader bar, while the upper suspension cabes support the spreader bar from the truck plate, which is hidden from view of the antenna beams by the reflector plate.  Both are visible in Figure \ref{payload}.\\
$^c$ Stainless steel flares surround the sky port.  See \S \ref{appconfig}.\\ 
\end{table*}

\section{Instrument peripherals contribution to measured radiometric temperature} \label{fltrain} 

The instrument core is in the far sidelobes of the antenna beams so its thermal emission to the radiometers is negligible.  However the flight train, consisting of the parachute, ladder, truck plate, FAA transmitter, and balloon are directly above the radiometer only 30$^\circ$ from the center of the beams.  Since the flight train is complicated and moves with the balloon rather than the gondola, a reflector constructed of aluminum foil covered foam board was attached to the gondola to hide these components from the antenna beams and instead reflect the sky into the antennas.  The V-shaped reflector shield and the spreader bar on which it is mounted can be seen in Figure \ref{payload}.  The total expected emission does not change much because of the presence of the reflector, but it is much easier to compute and more stable.  

We convolve the measured antenna pattern with the positions and emissivity estimates of the reflector plate, spreader bar, balloon, and several other components to estimate the radiometric temperature contribution from each in each band.  We consider both thermal emission from the components themselves and reflection of thermal emission from the 300 K ground.  These results are presented in Table \ref{tbl-3}.  We conservatively estimate an uncertainty of 30\% for these values.

Near the end of the flight the reflector was heated from 240 K to 300 K to look for the emission signal from the reflector shield in the radiometer outputs.  The predicted change due to the heating is smaller than the uncertainty.  No measurable signal was detected, putting a limit of the contribution from the reflector shield to $\sim$3 times the estimated signal. The geometric factor is highest for the 8 GHz radiometer as its beam boresight is closest to the reflector shield and spreader bar.

\section{Discussion}

The 2005 and 2006 flights have demonstrated the viability and utility of open-aperture cryogenic optics for absolute temperature microwave astrophysical measurements.  ARCADE 2 is able to maintain the external calibrator, antennas, and radiometers at temperatures near 2.7 K for many hours at 37 km altitude.  Cold boil-off gas reduces atmospheric condensation to negligible levels.  In the future, temperature gradients in the external calibrator can be greatly reduced by thermally standing off the aperture plate from its liquid helium tanks and controlling its temperature with SPID heaters.


\acknowledgments

 We thank the staff at CSBF for launch support.  We thank Victor Kulesh for contributions to the ground software, and Adam Bushmaker, Jane Cornett, Paul Cursey, Sarah Fixsen, Luke Lowe, and Alexandre Rischard for their work on the project.  We thank the Cryogenics Branch at GSFC for supporting the ARCADE 2 design and thermometer calibration, Todd Gaier for the 90 GHz amplifiers, and Custom Microwave, Bechdon, Inc., Flight Fab, and JMD for fabrication of components.  This research has been supported by NASA's Science Mission Directorate under the Astronomy and Physics Research and Analysis suborbital program. The research described in this paper was performed in part at the Jet Propulsion Laboratory, Californai Institute of Technology, under a contract with the National Aeronautics and Space Administration.  T.V. acknowledges support from CNPq grants 466184/00-0, 305219-2004-9 and 303637/2007-2-FA, and the technical support from Luiz Reitano.  C.A.W. acknowledges support from CNPq grant 307433/2004-8-FA.


\end{document}